\documentclass[11pt]{article}
\usepackage{amssymb}
\usepackage{amsmath, amsthm}
\newcommand{\be}{\begin{equation}}
\newcommand{\ee}{\end{equation}}
\begin{document}
\def\theequation{\arabic{section}.\arabic{equation}}
\begin{titlepage}
\title{$f(R)$ gravity: successes and challenges}
\author{Valerio Faraoni\\ \\
{\small \it Physics Department, Bishop's University}\\
{\small \it 2600 College St., Sherbrooke, Qu\'{e}bec, Canada 
J1M~1Z7}
}
\date{} \maketitle
\thispagestyle{empty}
\vspace*{1truecm}
\begin{abstract}
We review the state of the art of $f(R)$ theories of gravity (in their 
various formulations), which have been proposed as an 
explanation of the cosmic acceleration alternative to dark energy. The 
successes of $f(R)$ gravity are discussed, together with the challenges 
imposed by minimal criteria for their viability.
 \end{abstract} 
\vspace*{1truecm}
\begin{center}  
Presented at SIGRAV~2008, 18th Congress of the Italian Society of 
General Relativity and Gravitation, Cosenza, Italy September 22-25, 
2008. 
\end{center}
\setcounter{page}{1}
\end{titlepage}

\section{Introduction}
\setcounter{equation}{0}

The acceleration of the universe was discovered 
ten years ago using type Ia supernovae \cite{SN} and no definitive or 
truly   satisfactory 
explanation of this phenomenon has been given yet. This discovery has 
important implications not only for cosmology, but also for fundamental 
physics.  According to {\em WMAP} 
and the other experiments mapping anisotropies of the cosmic microwave 
background, if general relativity is the correct description 
of our 
universe, then approximately 76\% of its energy content  
is not dark or luminous matter, but is instead  
a mysterious form of {\em dark energy}, exotic, invisible, and  
unclustered.  Three main  classes of 
models for this cosmic 
acceleration have been proposed:
\begin{enumerate}
\item a  cosmological constant $\Lambda$
\item dark energy
\item modified gravity.
\end{enumerate} 
Naively, a cosmological constant  propelling the 
cosmic acceleration and eventually coming to dominate the universe, 
causing it to enter a  de Sitter phase without return, seems  
the most obvious explanation. However,  $\Lambda$  brings with it 
the notorious cosmological constant problem and the 
coincidence problem. To admit that $\Lambda$ is non-zero but still has 
the tiny 
value required to cause the current acceleration amounts to an extreme 
fine-tuning. It is not surprising, therefore, that most cosmologists 
reject this explanation, postulating that $\Lambda$ is exactly zero for 
reasons yet to be discovered (the cosmological constant problem is 
put in a drawer for the time being), and that a different  
explanation is to be found for the cosmic acceleration.

The second class of models, (mostly) within the context of general 
relativity,  postulates the existence of a dark energy fluid with 
equation of state $P \approx -\rho$ (where $\rho$ and $P$ are the 
energy density 
and pressure of the fluid, respectively), which comes to dominate late 
in the matter era. Dark energy could even  be phantom energy with 
equation of state such that $ P < -\rho $. Many dark energy models 
have been studied, none of which is totally convincing or free of 
fine-tuning problems, or can be 
demonstrated to be the ``correct'' one.

A third possibility consists in dispensing entirely with  
the mysterious dark energy and modifying  
gravity at the largest scales.\footnote{Historically, this approach 
was the correct one in explaining the precession of 
Mercury's perihelion: not due to an unseen mass but to Einstein's 
modification of Newtonian gravity.} Here we focus on modified gravity 
and, 
specifically, on the so-called $f(R)$ gravity theories.\footnote{We do 
not consider here braneworld models or other modifications of gravity 
that are sometimes also referred to as ``modified gravity''.}

$f(R)$ or ``modified'' gravity consists of infrared modifications  of 
general relativity that become important only at low curvatures,  late 
in the matter era. The Einstein-Hilbert 
action\footnote{Here $R$ is the Ricci curvature of the metric tensor 
$g_{ab} $ with 
metric determinant $g$, $\kappa \equiv 8\pi G$, $G$ is Newton's 
constant, and we follow the notations of Ref.~\cite{Wald}.} 
$ S_{EH}=\frac{1}{2\kappa}\int d^4x \, \sqrt{-g} \, 
R+S^{(matter)} $ 
is modified to 
\be \label{actionmetric}
S=\frac{1}{2\kappa} \int d^4x \, \sqrt{-g} \, f(R)+S^{(matter)} \;,
\ee
where $f(R)$ is a non-linear function of its argument \cite{CCT, CDTT}.

In principle, the metric tensor contains several degrees of freedom: 
tensor, vector, and scalar, massless or massive. In general relativity 
only the familiar massless spin~2 graviton propagates. When the 
Einstein-Hilbert action is modified, other degrees of freedom 
appear. The change $R-2\Lambda \rightarrow f(R)$ in the action brings to 
life, in addition to the massless graviton, a massive scalar mode which 
can drive the cosmic acceleration, and  will be discussed below. This 
is analogous to the inflaton  field driving  the 
accelerated expansion of the early universe, although at a much lower 
energy scale.

If terms quadratic in 
the Ricci and Riemann tensor, and possibly other curvature invariants, 
are included in the gravitational Lagrangian, $ f 
\left( R, R_{ab}R^{ab}, R_{abcd}R^{abcd}, \, ... \, \right)$,  
massive  gravitons and vector degrees of freedom appear. In the 
following we restrict ourselves to 
$f(R)$ theories of gravity in their various formulations, and we focus 
on their use as substitutes for dark energy;\footnote{$f(R)$ gravity, 
sometimes with an explicit coupling of matter to $R$ \cite{extraf}, has 
also been used as an alternative to galactic dark matter  
\cite{DM}.}  for  a more   
comprehensive discussion we refer the reader to \cite{review} and for 
short introductions to \cite{NojiriOdintsovreview, Straumann}.

We adopt a conservative point of view and regard $f(R)$ 
gravity more as a toy model than the correct theory of gravity, {\em 
i.e.}, we consider these theories as  a proof of 
principle that modifying gravity is a viable alternative to dark energy.  
However, we do not feel that one can claim that any of the $f(R)$ models 
proposed thus far  is the 
``correct'' one, or has exceptional support from the observational data. 
While it is true that many $f(R)$ models pass all the available 
experimental tests and fit the cosmological data, the same is true for 
many 
dark energy models, and it is currently impossible to use 
observational data to discriminate 
between most of them, and between dark energy and modified gravity 
models.

Modifying gravity is risky: unwanted consequences may be  violations of  
the experimental limits on the 
parametrized-post-Newtonian (PPN) parameters at terrestrial and 
Solar System scales \cite{Will}, 
instabilities, 
ghosts and, as in any newly proposed 
theory, the Cauchy problem could be ill-posed. These aspects are 
discussed in the next sections.

$f(R)$ gravity has  a long history: its origins can be loosely traced 
to Weyl's 1919 theory  in which a term quadratic in the Weyl tensor 
was added to the Einstein-Hilbert Lagrangian \cite{Weyl}. Later, $f(R)$ 
gravity  
received the attention  of many authors, including 
Eddington, Bach, Lanczos, Schr\"odinger, and Buchdahl. In the 1960's and 
1970's, it was found that  quadratic corrections to $S_{EH}$  were 
necessary to improve the 
renormalizability of general relativity \cite{renorma}, and in 
1980 quadratic 
corrections were found to fuel inflation without the need for scalar 
fields \cite{Starobinsky80}. Non-linear corrections are also motivated 
by string theories \cite{stringmotivations}. We refer the reader to 
\cite{Schmidt} for an historical review.

The prototype of $f(R)$ gravity \cite{CCT,CDTT} is the model
\be
f(R)=R-\mu^4/R \;,
\ee
where $\mu $ is a mass scale of the order of the present value of the 
Hubble parameter $\mu \sim H_0\sim 10^{-33}$~eV. Although ruled out 
by its weak-field limit  \cite{CSE} and by  
a violent instability \cite{DolgovKawasaki}, this model  gives the idea 
underlying  modified gravity:  the $ 1/R$ correction  
is negligible in comparison with $R$ at the high curvatures of the early 
universe, and kicks in only as 
$R\rightarrow 0$, late in the history of the universe.

Many forms of  the function $f(R)$ are found in the literature: here  
we discuss only general, model-independent, features of $f(R)$ gravity.

\section{The three versions of $f(R)$ gravity}
\setcounter{equation}{0}

Modified gravity comes in three versions:\\ 
1)~metric (or second order) formalism;\\
2)~Palatini (or first order) formalism; and\\
3)~metric-affine gravity.\\

\subsection{Metric $f(R)$ gravity}

In the metric formalism  \cite{CCT, CDTT}, the action 
is
\be  \label{metricaction2}
S_{metric}=\frac{1}{2\kappa}\int d^4 x \, \sqrt{-g} \, f(R)+S^{(matter)} 
\;.
\ee
Variation with respect to the (inverse) metric tensor $g^{ab}$ yields 
the field equation
\be \label{metricfieldeqs}
f'(R)R_{ab}-\frac{f(R)}{2} \, g_{ab}=\nabla_a\nabla_b f'(R) -g_{ab} \Box 
f'(R) +\kappa\, T_{ab} \;,
\ee
where a prime denotes differentiation with respect to $R$. The first two 
terms on the right hand side introduce 
fourth order derivatives of the metric, hence the name ``fourth 
order gravity'' sometimes given  to these theories.

The trace of eq.~(\ref{metricfieldeqs}) yields
\be \label{tracemetric}
3\Box f'(R)+Rf'(R)-2f(R)=\kappa \, T \;,
\ee
where $T\equiv {T^a}_a$ is the trace of the matter stress-energy tensor. 
This  second order differential equation for $f'(R)$  
differs deeply from the trace of the Einstein equation $R=-\kappa \, T$  
which, instead, relates algebraically the Ricci scalar to 
$T$. We already see that $f'(R)$ is indeed a dynamical variable, the  
scalar degree of freedom contained in the theory.

Formally, one can rewrite the field equation~(\ref{metricfieldeqs}) in 
the form of an effective Einstein equation as
\be
G_{ab}=\kappa \left( T_{ab}+T_{ab}^{(eff)} \right)
\ee
where
\be
T_{ab}^{(eff)}=\frac{1}{\kappa} \left[ \frac{ f(R)-Rf'(R)}{2}\, 
g_{ab}+\nabla_a \nabla_b f'(R)-g_{ab} \Box f'(R) \right] 
\ee
is an effective stress-energy tensor containing geometric terms. Of 
course, as 
usual when adopting this procedure, $T_{ab}^{(eff)}$ does not satisfy 
any energy condition and the effective energy density is, in general,  
not positive-definite. As is clear from these equations, in $f(R)$ 
gravity one can define an effective gravitational coupling 
$G_{eff}\equiv G/f'(R)$  in a way analogous to what is done in 
scalar-tensor theories. Hence,   $f'(R)$ must be positive in order for 
the  graviton to carry  positive kinetic energy.

In the spatially flat Friedmann-Lemaitre-Robertson-Walker (FLRW) metric  
 adopted as the kinematic description of our universe,
\be
ds^2=-dt^2 +a^2(t) \left( dx^2+dy^2+dz^2 \right) \;,
\ee
the field equations of metric $f(R)$ cosmology assume the form
\begin{eqnarray}
& & H^2=\frac{\kappa}{3f'(R)} \left[ 
\rho^{(matter)}+\frac{Rf'(R)-f(R)}{2}-3H \dot{R} f''(R) \right] \;,\\
&&\nonumber \\
&& 2\dot{H}+3H^2= -\frac{\kappa}{f'(R)} \left[ 
P^{(matter)}+ f'''(R) \left( \dot{R}\right)^2 +2H\dot{R} 
f''(R)+\ddot{R}f''(R) \right. \nonumber \\
&& \nonumber \\
&& \left. +\frac{ f(R)-Rf'(R) }{2} \right] \;,
\end{eqnarray}
where an overdot denotes differentiation with respect to the comoving 
time $t$. The corresponding phase space is a 2-dimensional curved 
manifold embedded in a 3-dimensional space and with a rather complicated 
structure  \cite{deSouzaFaraoni}.

\subsection{Palatini $f(R)$ gravity}

In the Palatini approach, both the metric $g_{ab}$ and the connection 
$\Gamma^a_{bc}$ are independent variables, {\em i.e.}, the 
connection is not the metric connection of $g_{ab}$. While in general 
relativity the metric and Palatini variations produce the same 
(Einstein) equations, this is no longer true for  
 non-linear Lagrangians.\footnote{The requirement that the Palatini and 
metric variations give the  same field equations selects Lovelock 
gravity \cite{Exirifard}, of which general relativity is a special case.} 

Shortly after metric $f(R)$ theories were 
proposed as alternatives to dark energy, also the Palatini version 
was adanced for the same purpose, originally in its $f(R)=R-\mu^4/R$ 
incarnation  
\cite{Vollick}. The Palatini action is
\be\label{actionPalatini}
S_{Palatini}=\frac{1}{2\kappa}\int d^4 x \, \sqrt{-g} \, f\left( 
\tilde{R} 
\right) +S^{(matter)}\left[ g_{ab}, \psi^{(m)} \right] \;.
\ee
There are two Ricci 
tensors: $R_{ab}$, which is constructed using the metric connection of 
the (unique) 
physical metric $g_{ab}$, and $\tilde{R}_{ab}$ which is the Ricci 
tensor of the non-metric connection $\Gamma^a_{bc}$. The latter gives 
rise to the scalar $\tilde{R}\equiv g^{ab}\tilde{R}_{ab}$. 
The matter part of the action does not depend explicitly from the 
connection $\Gamma$, but only from the metric and the matter fields, 
collectively denoted with $\psi^{(m)}$.

Variation of the Palatini action~(\ref{actionPalatini}) yields the field 
equation
\be \label{Palatinifieldeq1}
f'(\tilde{R}) \tilde{R}_{ab}-\frac{ f(\tilde{R})}{2} \, g_{ab}=\kappa \, 
T_{ab} \;.
\ee
Note the absence of second covariant derivatives of $f'$, in contrast 
with eq.~(\ref{metricfieldeqs}). Variation with respect to the 
independent connection produces the field equation
\be \label{Palatinifieldeq2}
\tilde{\nabla}_d \left( \sqrt{-g} \, f'(\tilde{R}) g^{ab} 
\right)-\tilde{\nabla}_d \left( \sqrt{-g} \, f'(\tilde{R}) 
g^{d(a}\right) \delta^{b)}_c =0 \;,
\ee
where $\tilde{\nabla}_c$ denotes the covariant derivative associated to 
this non-metric connection $\Gamma$.
The trace of eqs.~(\ref{Palatinifieldeq1}) and (\ref{Palatinifieldeq2}) 
yields
\be \label{Palatinitrace}
f'(\tilde{R}) \tilde{R} -2f(\tilde{R})=\kappa \, T 
\ee
and
\be \label{Palatinifieldeq3}
\tilde{\nabla}_c \left( \sqrt{-g} \, f'(\tilde{R}) g^{ab} \right)=0 \;.
\ee
Eq.~(\ref{Palatinifieldeq3}) tells us that $\tilde{\nabla}_c$ is the 
covariant derivative of the metric
\be
\tilde{g}_{ab}\equiv f'( \tilde{R}) g_{ab} 
\ee
conformally related to $g_{ab}$.  Note that eq.~(\ref{Palatinitrace}) is 
an algebraic 
(or trascendental, depending on the form of the function $f$) and not a 
differential equation for $f'(\tilde{R})$: hence, this quantity 
is non-dynamical, contrary to metric $f(R)$ gravity. 
This lack of dynamics has important consequences explored in 
the following sections.
It is possible to eliminate completely the non-metric connection from 
the field equations, which are then rewritten as
\begin{eqnarray}
&& G_{ab}=\frac{\kappa}{f'}\, T_{ab}-\frac{1}{2}\left( 
R-\frac{f}{f'}\right) g_{ab} +\frac{1}{f'} 
\left(\nabla_a\nabla_b 
-g_{ab}\Box \right) f'\nonumber\\
&&\nonumber \\
&& -\frac{3}{2(f')^2}  \left[  
\nabla_a f' \nabla_b 
f' -\frac{1}{2} g_{ab} \nabla_c f' \nabla^c f' \right] \;.
\label{Palatinireformulated}
\end{eqnarray}

\subsection{Metric-affine $f(R)$ gravity}

In metric-affine $f(R) $ gravity \cite{metricaffine}, also the matter 
part of the action 
\be
S_{affine}=\frac{1}{2\kappa}\int d^4 x \, \sqrt{-g} \, f\left( \tilde{R} 
\right) +S^{(matter)}\left[ g_{ab}, \Gamma^a_{bc}, \psi^{(m)} \right] 
\;,
\ee
depends explicitly on the  connection 
$\Gamma$, which is possibly non-symmetric. This leads to a torsion 
associated with matter, and to a modern revival of 
torsion theories. These were originally introduced within a 
non-cosmological context, with the spin of elementary particles 
coupling to the torsion. Metric-affine $f(R)$ gravity has not yet been 
explored in great detail, especially with respect to its cosmological 
consequences. For these reasons, in the following we focus on metric 
and Palatini $f(R)$ gravity.

\section{Equivalence of metric and Palatini $f(R)$ gravities with 
Brans-Dicke theories}
\setcounter{equation}{0}

If $f''(R) \neq 0$, metric modified gravity is equivalent to an 
$\omega=0$ Brans-Dicke theory\footnote{The general form of 
the Brans-Dicke 
action is $ S_{BD} = \frac{1}{2\kappa} \int d^4x \, \sqrt{-g} \left[ 
\phi R -\frac{\omega}{\phi} \, \nabla^c\phi \nabla_c\phi -V(\phi) 
\right] +S^{(matter)} $.} \cite{BD}, while Palatini modified gravity 
is 
equivalent to an $\omega=-3/2$ one.  This equivalence has been proposed 
and rediscovered, for particular theories or in general, many 
times over the years \cite{STequivalence}.

\subsection{Metric formalism}

Assuming that $f''(R) \neq 0$ and beginning with the 
action~(\ref{actionmetric}), one introduces the 
auxiliary scalar field $\phi =R$ and considers the action
\be \label{equivalentmetric}
S=\frac{1}{2\kappa} \int d^4 x \, \sqrt{-g} \left[ \psi( \phi)R -V(\phi) 
\right] +S^{(matter)} \;,
\ee
where 
\be
\psi(\phi) = f'(\phi) \;, \;\;\;\;\;\;
V(\phi)=\phi f'(\phi)-f(\phi) \;.
\ee

The action~(\ref{equivalentmetric}) trivially  reduces 
to~(\ref{actionmetric})   for metric $f(R)$ gravity if $\phi=R$. 
Vice-versa, the variation 
of~(\ref{equivalentmetric}) with respect to $g^{ab}$ gives
\be
G_{ab}=\frac{1}{\psi}\left( \nabla_a\nabla_b \psi-  g_{ab}\Box\psi 
-\frac{V}{2}\, g_{ab} \right)+\frac{\kappa}{\psi} \, T_{ab} \;,
\ee
while varying with respect to $\phi$ yields
\be
R\, \frac{d\psi}{d\phi} -\frac{dV}{d\phi}=\left( R-\phi 
\right)f''(\phi)=0
\ee
and $\phi=R$ under the assumption $f''\neq 0$. Hence, the scalar 
field  $\phi=R$  is dynamical and 
satisfies the trace equation
\be
3f''(\phi)\Box \phi+3f'''(\phi)\nabla^c\phi\nabla_c \phi +\phi 
f'(\phi)  -2f(\phi) =\kappa \, T\;.
\ee
This scalar is massive: as discussed later, the analysis of small 
perturbations of de Sitter space allows one to compute explicitly its  
mass squared 
\be
m_{\phi}^2=\frac{1}{3} \left( \frac{f_0'}{f_0 ''}-R_0 \right) \;,
\ee
where a zero subscript denotes  quantities evaluated at the constant 
curvature of the de Sitter background.  
It turns out to be more convenient to consider the scalar 
$ \psi \equiv f'(\phi)$, which satisfies
\be
3\Box \psi +2 U(\psi) -\psi\, \frac{dU}{d\psi}=\kappa \, T 
\ee
with $ U(\psi)=V(\phi(\psi))-f(\phi(\psi)) $.
It is clear, therefore, that the theory contains a scalar degree of 
freedom, and the action
\be
S=\frac{1}{2\kappa} \int d^4 x \, \sqrt{-g} \left[ \psi R -U(\psi) 
\right] +S^{(matter)} \;,
\ee
 is recognized as an 
$\omega=0$ Brans-Dicke theory. This theory, called ``massive dilaton 
gravity'' was originally introduced in the 1970's in order to generate a 
Yukawa term in the Newtonian limit  \cite{OHanlon72}. The assumption 
$f''\neq 0$ can  
be seen as the requirement that the change of variable $R\rightarrow 
\psi(R)$ be invertible.

\subsection{Palatini formalism}

In the Palatini case, the discussion of the equivalence with a 
Brans-Dicke theory proceeds in a way analogous to 
that of the metric formalism. One begins with the 
action~(\ref{actionPalatini}) and introduces  
$\phi  =\tilde{R}$ and $\psi \equiv f'(\phi)$. Then, apart from a 
boundary term  that can be neglected for classical purposes, the action 
is rewritten, in terms of the metric $g_{ab}$ and of its Ricci 
tensor $R_{ab}$, as
\be \label{equivalentPalatini}
S_{Palatini}=\frac{1}{2\kappa}\int d^4x \, \sqrt{-g} \left[ \psi R 
+\frac{3}{2\psi} \, \nabla^c \psi\nabla_c \psi -V(\psi) \right] 
+S^{(matter)} \;,
\ee
where we used  the fact that, since $\tilde{g}_{ab}=\psi \, g_{ab}$, the 
Ricci curvatures of  $g_{ab}$ and 
$\tilde{g}_{ab}$ are related by  
\be
\tilde{R}=R +\frac{3}{2\psi} 
\nabla^c\psi \nabla_c\psi-\frac{3}{2} \Box \psi \;.
\ee 
The  action~(\ref{equivalentPalatini}) is recognized as a Brans-Dicke 
theory with Brans-Dicke parameter $\omega=-3/2$.

\section{Criteria for viability}
\setcounter{equation}{0}

In order for $f(R)$ gravity to be successful, it is not sufficient that 
it serves the purpose for which it was introduced in the cosmological 
context, but it must also pass the tests imposed by Solar System and 
terrestrial experiments on relativistic gravity, and it must satisfy 
certain minimal criteria for viability. Overall, these are:

\begin{itemize}

\item possess the correct cosmological dynamics;
\item not suffer from instabilities and ghosts;
\item have the correct Newtonian and post-Newtonian limit;
\item give rise to cosmological perturbations compatible with the data 
from the cosmic microwave background and large scale structure surveys; 
and
\item have  a well-posed Cauchy problem.
\end{itemize}

The failure to satisfy even a  single one of these criteria is 
taken as a statement that the theory is doomed. These viability criteria 
are examined in the following.

\subsection{Correct cosmological dynamics}

In the opinion of  most cosmologists, in order to be acceptable a 
cosmological model must exhibit  early inflation (or an alternative way 
to solve the horizon, flatness, and monopole problem together with a 
mechanism to 
generate density 
perturbations), followed by a radiation-dominated era and 
a matter-dominated era, and then by the present accelerated epoch that 
$f(R)$ theories were resurrected  to explain.  The future era is usually 
found to be an eternal  
de Sitter attractor phase, or a Big Rip singularity truncating the 
history of the 
universe at a finite time.

Smooth transitions between  different eras are required. It has 
been pointed out that the exit 
from the radiation era, in particular, may have problems in many models  
\cite{Amendolaetal}, a warning 
that care must be exerted in building $f(R)$ cosmologies. Ultimately, 
exit from the radiation, or any 
era can be achieved. Take, for example, what we could name 
``designer $f(R)$ gravity'': one can prescribe a desired expansion 
history of the universe by a choice of the scale factor $a(t)$ and 
then integrate an ODE that determines  the function $f(R)$ that produces 
$a(t)$ \cite{designerf(R)}. 
In general, this function is not unique and assumes  
rather contrived forms (not the usual $R-\mu^{2(n+1)}/R^n$, or simple 
forms like 
that).

\subsection{Instabilities}

The prototype model in the  discussion of instabilities is again the 
choice 
$f(R)=R-\mu^4/R$ with $\mu\sim H_0\sim 10^{-33}$~eV.  Shortly 
after it was proposed, this model was found to suffer from a 
catastrophic  (``Dolgov-Kawasaki'') instability  
\cite{DolgovKawasaki}. The stability analysis  was later 
generalized to any metric $f(R)$ theory \cite{mattmodgrav} and the 
extension to even more general gravitational theories has been pursued 
\cite{Zerbini}. One 
proceeds  by parametrizing the deviations from general relativity as
\be
f(R)=R+\epsilon \varphi(R) \;,
\ee
where $\epsilon$ is a small positive constant with the dimensions of  
a mass 
squared and the function $\varphi$ is dimensionless. The trace equation 
for the Ricci scalar $R$ takes the form
\be 
\Box R+\frac{\varphi '''}{\varphi ''} \, \nabla^c R \nabla_c R +\left( 
\frac{\epsilon \varphi ' -1}{3\epsilon \varphi ''} \right) R 
=\frac{\kappa \, T}{3\epsilon \varphi ''}+\frac{2\varphi}{3\varphi ''} 
\;.
\ee
Next, one expands around a de Sitter background and   
writes the metric {\em locally} as
\be \label{localdS}
g_{ab}=\eta_{ab}+h_{ab} \;,
\ee
while the scalar degree of freedom $R$ is expanded as
\be 
R=-\kappa\, T +R_1 \;,
\ee
with $R_1$ a perturbation. To first order, the trace equation yields the 
dynamical equation for $R_1$
\be
\ddot{R}_1 -\nabla^2 R_1 -\frac{2\kappa \varphi '''}{\varphi ''} \, 
\dot{T}\dot{R}_1+\frac{2\kappa \varphi '''}{\varphi ''} \, \vec{\nabla}T 
\cdot \vec{\nabla}R_1 + \frac{1}{3\varphi ''}\left( 
\frac{1}{\epsilon}-\varphi ' \right) R_1=\kappa \, \ddot{T}-\kappa 
\nabla^2 T -\frac{ \left( \kappa T \varphi^2 +2\varphi \right)}{3\varphi 
''} \;.
\ee
The last term on the left hand side is dominated by the term in 
$\epsilon^{-1}$ and gives the effective mass squared of $R_1$
\be
m^2 \simeq \frac{1}{3\epsilon \varphi ''} \;,
\ee
from which one deduces that the theory is 

\begin{itemize} 
\item {\em stable} if $f''(R)>0$
\item   {\em unstable} if $f''(R)<0 \;.$
\end{itemize}
The case of general relativity 
is excluded by the assumption $f''\neq 0$, but the well-known stability 
in this case allows one to extend the stability criterion for metric 
$f(R)$ gravity to be $f'' \geq 0$. 

As an example, the prototype model $f(R)=R-\mu^4/R$, which has $f''<0$ 
is unstable. The time scale for the onset of this instability is 
dictated by the smallness of the scale $\mu$ and is seen to correspond 
to 
$\sim 10^{-26}$~s \cite{DolgovKawasaki}, making this an  
explosive instability.

One can give  a physical interpretation of this result 
as follows \cite{myinterpretation}: remembering that the effective 
gravitational coupling is $G_{eff}=G/f'(R)$, if 
$dG_{eff}/dR=-f''G/(f')^2>0$ (which corresponds to $f''<0$), then 
$G_{eff}$ increases with $R$ and a 
large curvature causes gravity to become stronger, which in turn causes 
a larger $R$, in a positive feedback mechanism  driving the system 
away. If instead $dG_{eff}/dR<0$, then  
a negative feedback   damps the increase in the  
gravitational coupling strength.

Palatini $f(R)$ gravity, by contrast, is described by second order 
field equations, the trace equation $ f'( \tilde{R}) \tilde{R} 
-2f(\tilde{R})=\kappa \, T$  is  not a differential equation but rather
a non-dynamical algebraic one and, therefore, there is no 
Dolgov-Kawasaki instability  
\cite{SotiriouPalatiniinstab}.

The previous analysis for metric $f(R)$ gravity obtained with the 
local 
expansion~(\ref{localdS}) 
is necessarily limited to short wavelengths (compared to  the 
curvature radius), but can be extended to the longest wavelenghts 
\cite{mydS}. This is necessarily more complicated because these modes 
suffer from the notorious gauge-dependence problems of cosmological 
perturbations and  
a covariant and gauge-invariant formalism is needed. 
We assume that the background space is de Sitter and consider the  
general action
\be
S=\int d^4 x \, \sqrt{-g} \, \left[ \frac{f \left(\phi, R \right)}{2} 
-\frac{\omega(\phi)}{2}\, \nabla^c \phi \nabla_c \phi -V(\phi) \right] 
\;,
\ee
which contains $f(R)$ and  scalar-tensor gravity, and mixtures of them. 
On a FLRW background, the field equations become
\begin{eqnarray}
& & H^2= \frac{1}{3f'} 
\left(\frac{\omega}{2}\,\dot{\phi}^2+\frac{Rf'-f}{2}+V-3H\dot{f}\right) 
\;, \\
&&\nonumber \\
&& \dot{H}=\frac{-1}{2f'} \left( \omega \dot{\phi}^2 
+ \ddot{f'}-H\dot{f'} \right)  \;, \\
&&\nonumber \\
&& \ddot{\phi}+3H\dot{\phi} +\frac{1}{2\omega}\left( 
\frac{d\omega}{d\phi} 
\dot{\phi}^2 -\frac{\partial f}{\partial \phi} +2 \, \frac{dV}{d\phi} 
\right)=0 \;.
\end{eqnarray}
de Sitter space is  a solution  subject to the conditions
\be
6H_0^2 f_0'-f_0 +2V_0=0 \;, \;\;\;\;\;\;\;\;
f_0'=2V_0' \;.
\ee
An analysis \cite{mydS} using the covariant and gauge-invariant 
Bardeen-Ellis-Bruni-Hwang  
formalism \cite{Bardeen} in the version given by Hwang \cite{Hwang} for 
alternative 
gravitational theories yields the stability condition of de Sitter space 
in metric $f(R)$~gravity with respect to {\em inhomogeneous} 
perturbations
\be \label{stabilitydS}
\frac{ (f_0')^2 -2f_0 f_0''}{f_0' f_0''} \geq 0 \;,
\ee
which is obtained in the zero momentum limit. This condition coincides 
with the stability condition with respect to {\em homogeneous} 
perturbations \cite{myinterpretation}.

At this point it is worth checking that the equivalence between metric 
$f(R)$ gravity and an $\omega=0$ Brans-Dicke theory holds also at the 
level of perturbations; previous doubts to this regard  
\cite{Faraonidoubts, OdintsovNojiridoubts} have now been dissipated.

For the   $\omega=0$ Brans-Dicke theory, the stability condition of de 
Sitter 
space with respect to inhomogeneous perturbations is given again by 
eq.~(\ref{stabilitydS}), while that for stability with respect 
to  homogeneous perturbations is
\be
\frac{ (f_0')^2 -2f_0 f_0''}{f_0'} \geq 0 \;. 
\ee
This is again equivalent to~(\ref{stabilitydS}) provided that stability 
against {\em local}  perturbations, expressed by 
$f_0''>0$, is assumed. Therefore, there is complete equivalence between 
metric $f(R)$ gravity and $\omega=0$ Brans-Dicke theory also at the 
level of cosmological perturbations.

Going beyond the linear approximation, metric $f(R)$ theories have been 
found to be susceptible to another, non-linear, instability, which makes 
it hard to build models of relativistic stars in strong gravity. In 
these situations, a singularity develops for large $R$, which was 
discovered in \cite{Frolovetc}. Although 
this problem needs further study, it seems that, in order to avoid 
this singularity requires some degree of fine-tuning. At present, this 
is probably the biggest challenge for metric $f(R)$ theories.

\subsection{Ghosts}

Ghosts are massive states of negative norm which cause lack of unitarity 
and are common when trying to generalize Einstein's gravity. The 
good news here are that $f(R)$ gravity is ghost-free. More general 
theories of the form $f\left( R, R_{ab}R^{ab}, R_{abcd} R^{abcd}, ... 
\right) $, in general, contain ghost fields. A possible exception 
(under certain conditions \cite{GBghosts}) is the case 
in which the extra terms appear in the Gauss-Bonnet combination ${\cal 
G}=R^2-4R_{ab}R^{ab}+R_{abcd}R^{abcd}$, as in $f=f\left( R, {\cal G} 
\right)$ In this case, the field equations are of second order and 
there are no ghosts \cite{NunezSolganik04, Comelli05, 
NavarroVanAcoleyen06}.

\subsection{Weak-field limit (metric formalism)}

Early work on the weak-field limit of both metric and Palatini $f(R)$ 
gravity was subject to errors and incompleteness (see \cite{review} for 
details\footnote{The limit of $f(R)$ gravity to general relativity has 
the character of a singular limit, and even the limit to general 
relativity of Brans-Dicke theory is not free from ambiguities 
\cite{BDlimit}.}); a satisfactory 
treatment for the prototype model $f(R)=R-\mu^4/R$ in the metric 
formalism was given in  \cite{Chibaetal06} and then generalized 
to arbitrary forms of  the function $f(R)$ in \cite{Olmo07, CSE}.

In order to assess whether the limits set on the PPN parameter $\gamma$  
by the 
available Solar System experiments, one needs to find the weak-field 
solution of the field equations and compute this parameter. One 
considers a static, spherically symmetric, non-compact body which 
constitutes a perturbation of a  
background de Sitter universe. The line element is written as
\be
ds^2=-\left[ 1+2\Psi(r)-H_0^2r^2 \right] dt^2 +\left[ 1+2 \Phi(r) +H_0^2 
r^2 \right] dr^2 +r^2 d\Omega^2 
\ee
in Schwarzschild coordinates, where $d\Omega^2$ is the line element on 
the unit 2-sphere and $\Psi$ and $\Phi$ are post-Newtonian potentials. 
These are of small amplitude, $\left|\Psi (r) \right| , 
\left|\Phi (r) \right| <<1$, and one considers small (non-cosmological) 
scales so that $H_0r <<1$, while expanding the Ricci scalar around the 
constant curvature of the background de Sitter space, $R(r)=R_0+R_1$. 
The PPN parameter $\gamma$ is given by $
\gamma =-\Phi(r) / \Psi(r) $ \cite{Will}. Three assumptions are made 
\cite{CSE}:

\begin{enumerate}
\item $f(R)$ is analytical at $R_0$;
\item $mr<<1$, where $m$ is the effective mass of the scalar degree of 
freedom of the theory, {\em i.e.}, it is assumed that this scalar field 
is light and has a range larger than the size of the Solar System (we 
remind the reader that there are no experimental constraints of scalars 
with range $m^{-1} <0.2$~mm).
\item For the matter composing the spherical body, the pressure is 
negligible, $P\simeq 0$, so that $ T=T_0+T_1 \simeq -\rho$.
\end{enumerate}

The first and the last assumption are not stringent, but the second one 
is, as will be clear below. The trace equation~(\ref{tracemetric}) 
yields 
the equation for the Ricci scalar perturbation 
\be
\nabla^2 R_1 -m^2 R_1 =\frac{-\kappa \rho}{3f_0''} \;,
\ee
where 
\be \label{cicci}
m^2 =\frac{  (f_0')^2 -2f_0f_0''}{3f_0'  f_0''} 
\ee
is the effective mass squared of the scalar. Eq.~(\ref{cicci}) coincides 
with the expression obtained in the gauge-invariant stability analysis 
of de Sitter space and in propagator calculations.

If $mr<<1$, the solution of the linearized field equations is
\begin{eqnarray}
&& \Psi(r)=\frac{-\kappa M}{6\pi f_0'}\, \frac{1}{r} \;, \\
&&\nonumber \\
&& \Phi(r) =\frac{\kappa M}{12 \pi f_0'} \, \frac{1}{r} \;.
\end{eqnarray}
The PPN parameter sought for is, therefore,
\be
\gamma =\frac{-\Phi(r)}{\Psi(r)}=\frac{1}{2} \;,
\ee
in gross violation of the (recently improved) experimental limit 
\cite{BertottiIessTortora}
\be
\left| \gamma -1 \right| <2.3 \cdot 10^{-5} \;.
\ee
This result would be the end of metric $f(R)$ gravity if the 
assumptions made in the calculation were satisfied. However, this is 
not 
the case for assumption~2): $mr$ is not always less than unity due to 
the 
{\em chameleon effect}. This consists in the effective mass  $m$ 
depending on the curvature or, alternatively, the matter density of the 
environment. The scalar degree of freedom can be 
short-ranged (say $m > 10^{-3}$~eV, corresponding to a range $\lambda 
< 0.2$~mm) at 
Solar System densities and evade the experimental constraints, while 
being long-ranged at cosmological densities and thus being able to 
affect the 
cosmological dynamics \cite{NavarroVanAcoleyen06, Faulkneretal06}. 
Although at a first glance the chameleon effect could 
be seen as a contrived and  fine-tuned   mechanism, $f(R)$ gravity is 
rather complicated and the effective range does indeed depend on the 
environment. The chameleon mechanism is well-known and 
accepted in  quintessence models, in which it was 
discovered for the scalar 
field potential $V(\phi) \approx 1/\phi$ \cite{chameleon}. Many forms 
of the function $f(R)$ are known to exhibit the chameleon mechanism and 
pass the observational tests. For example, the model
\be \label{Faulknermodel}
f(R)=R-\left(1-n \right) \mu^2 \left( \frac{R}{\mu^2} \right)^n 
\ee
is compatible with the PPN limits if $ \mu \sim 
10^{-50}$~eV$\sim10^{-17} H_0$ \cite{Faulkneretal06}. It is obvious
that a correction term  $\sim R^n$ with $n<1$ to the Einstein-Hilbert 
Lagrangian $R$  will come to dominate as $R\rightarrow 0^{+}$ (for 
example, 
for $n=1/2$, $\sqrt{R} >R $ as $R\rightarrow 0$). The 
model~(\ref{Faulknermodel}) is compatible with the experimental data but 
it could be essentially indistinguishable from a dark energy  model. 
Hope of discriminating between dark energy and $f(R)$ models, or between 
different modified gravities relies on the study of the 
growth of cosmological 
perturbations.

\subsection{Correct dynamics of cosmological perturbations}

The expansion history of the universe alone is not sufficient to 
discriminate between various models, but the growth of structures 
depends 
on the theory of gravity and has the potential to achieve this goal. 
Song, Hu, and Sawicki \cite{SongHuSawicki06} assumed an expansion 
history $a(t)$ typical of the $\Lambda$CDM model and found that vector 
and tensor modes are not affected by $f(R)$ corrections to Einstein 
gravity, to lowest order, while scalar modes are. They also found the 
condition $f''(R)>0$ for  the stability of scalar perturbations, in 
agreement with the arguments discussed above. The most interesting 
results are that $f(R)$ corrections lower the large angle anisotropies 
of the cosmic microwave background and produce  correlations 
between cosmic microwave background and 
galaxy surveys different from those of  dark energy models.

Overall, the study of structure formation in modified gravity is still 
work in progress, and often is performed within the context of specific 
models, 
some of which are already in trouble because they do not pass the 
weak-field limit or the stability constraints. A similar situation 
holds for all Palatini $f(R)$ models, and for this reason, their 
weak-field limit and cosmological perturbations are not discussed 
here.

\subsection{The Cauchy problem}

A physical theory must have predictive power and, therefore, 
a  well-posed  initial value problem.   General relativity satisfies 
this 
requirement for  ``reasonable'' forms of matter \cite{Wald}. The 
well-posedness of 
the Cauchy problem for vacuum $f(R)$ gravity was briefly discussed for 
special metric models  in earlier papers 
\cite{Noakes}.  Thanks to the  
equivalence between $f(R)$ gravity and 
scalar-tensor theory when 
$f''(R)\neq 0$,  the   Cauchy 
problem can be reduced to the  analogous one  for 
Brans-Dicke gravity with $\omega=0, -3/2$.
 That the  
initial value  problem  is well-posed was demonstrated for 
particular scalar-tensor theories 
in \cite{Cocke, Noakes} and a general analysis has   
only recently been performed \cite{Salgado, 
Salgado2}. This work, 
however, does not cover  the $\omega=0, -3/2 $ cases.

A system of $3+1$ equations of motion is {\em well-formulated}
if it can be written as a 
system of equations of only first order  in 
both time and space derivatives. If the latter is cast in 
the full  first order form 
\begin{equation}
\partial_t \, \vec{u} + M^i \nabla_i \vec{u}=\vec{S}\left( 
\vec{u}\right) ,
\end{equation}
where $\vec{u}$ collectively denotes the fundamental variables 
$h_{ij}, K_{ij}$, {\em etc.} introduced below, $M^i$ is 
called the {\em characteristic matrix} of the system, and 
$\vec{S}\left(  \vec{u} \right)$ describes source terms and 
contains only the fundamental variables but not their 
derivatives. The initial 
value formulation is {\em well-posed} if the system of PDEs is {\em 
symmetric hyperbolic} ({\em  
i.e.}, the matrices $M^i$ are symmetric)  and {\em strongly 
hyperbolic} if $ s_iM^i$ has a real set of 
eigenvalues and a complete 
set of eigenvectors for  any 1-form $s_i$, and obeys some 
boundedness conditions \cite{Solin}. 

In short, the result obtained in \cite{TremblayFaraoni} is that the 
Cauchy problem 
for metric $f(R)$ gravity is 
well-formulated   and is well-posed in vacuo and with 
``reasonable'' forms of matter ({\em i.e.}, perfect fluids, scalar 
fields, or the Maxwell field) while  for  Palatini 
$f(R)$ gravity,  instead, the  Cauchy problem is  
not well-formulated nor well-posed due to the presence of 
higher derivatives of  the 
matter  fields in the field equations and to the fact that 
it is impossible to eliminate them \cite{TremblayFaraoni}.

Let us consider the  scalar-tensor action 
\begin{equation}
S = \int d^{4}x\sqrt{-g}\left[ \frac{f(\phi) R}{2\kappa}  - 
\frac{1}{2}\nabla^{c}\phi\nabla_{c}\phi  - 
V(\phi)\right] + S^{(matter)} \;;
\end{equation}
Salgado  \cite{Salgado}  
showed that the corresponding Cauchy  problem is 
well-posed 
in vacuo and  well-formulated 
otherwise.  With the exception of $\omega=-3/2$, Salgado's 
results can be extended to the more 
general action  
\begin{equation}
 S \!=\! \int \! d^{4}x\sqrt{-g}\left[ \frac{f(\phi) R}{2\kappa} - 
\frac{\omega
(\phi)}{2}\nabla^{c} \phi \nabla_{c}\phi - V(\phi)\right] \!+\!  
S^{(matter)} \;,
\end{equation}
which contains the additional coupling function $\omega(\phi)$ 
\cite{TremblayFaraoni}. 

Setting $\kappa = 1$ in this section and performing a  $3+1$ 
Arnowitt-Deser-Misner  
splitting, one introduces 
 lapse $N$, shift $N^i$, spatial metric $h_{ij}$,  extrinsic curvature 
$K_{ij}$, and spatial gradient $Q_i$ of $\phi$ 
\cite{Wald, Reula,  Salgado}. Assume that a 
time 
function $t$ 
exists such that the spacetime $\left( M, g_{ab} \right)$ admits a  
foliation with  hypersurfaces $\Sigma_{t}$ of constant $t$ with 
unit timelike 
normal $n^{a}$.  The 
3-metric and projection operator on $\Sigma_t$ are $h_{ab} = 
g_{ab}+n_{a}n_{b}$ and  ${h^{a}}_{b}$, 
respectively, with 
\begin{eqnarray}
n^{c}n_{c} = -1 , \,\,\,\, h_{ab}n^{b} = 
h_{ab} n^{a} = 0 , \,\,\,\, {h_{a}}^{b} 
h_{bc} = h_{ac} \;, \noindent \\
\label{nolabel}
\end{eqnarray}
\begin{equation}
ds^{2} = -\left( N^{2} - N^{i}N_{i} \right) dt^{2} - 
2N_{i}dtdx^{i} + h_{ij}dx^{i}dx^{j} 
\end{equation}
$\left( i,j = 1,2,3 \right)$,
while  $ Q_{c} \equiv  D_{c} \phi $, 
and the momentum of $\phi $ is $
\Pi = {\cal L}_n\phi = n^{c}\nabla_{c}\phi $. Moreover, 
\begin{eqnarray}
\!\!K_{ij} \!\!\!&=&\!\!\! -\nabla_i n_j \!= - \frac{1}{2N}\! 
\left(\frac{\partial h_{ij} }{\partial t}  + 
D_i N_j 
+ D_j N_i \right) ,\\
&&\qquad\Pi = \frac{1}{N}\left(\partial_t\phi+N^{c} 
Q_{c}\right) ,\\
&& \partial_t Q_i + 
N^l\partial_l Q_i +Q_l \partial_iN^l=D_i\left(N\Pi\right) .
\end{eqnarray}
Omitting the calculations, the reduced $3+1$ equations 
 are \cite{TremblayFaraoni}
\begin{eqnarray}
&& \partial_t {K^i}_j + N^l\partial_l {K^i}_j + 
{K^i}_l\partial_j N^l - 
{K_j}^l\partial_l  N^i 
+ D^i D_j N \nonumber\\
&& \quad- ^{(3)}{R^i}_j N - NK{K^i}_j + \frac{N}{2\phi}\delta^i_j 
\left[ 2V(\phi) +  \square \phi \right]  \nonumber \\
&& \quad+ \frac{N}{\phi} \left( D^i Q_j + \Pi {K^i}_j  \right) 
+ \frac{N\omega_0}{\phi^2}Q^i Q_j \nonumber 
\\
&&\qquad = \frac{N}{2\phi}\left[ 
\left( S^{(m) } - E^{(m)} \right) \delta^i_j - 2 {S^{(m) \,\, 
i}}_j \right] \;,\\
&& \partial_t K + N^l\partial_l K + ^{(3)}\Delta N - 
NK_{ij}K^{ij}\nonumber \\
&& \quad - \frac{N}{\phi}\left( D^{\nu}Q_{\nu} + 
\Pi K \right) - \frac{\omega_{0}N}{\phi^2}\Pi^2 
\nonumber \\
&&  \qquad
 = \frac{N}{2\phi}\left[ -2V(\phi) - 3\square\phi + S^{(m)} + 
E^{(m)} 
\right] \;,\label{CP74}
\end{eqnarray}
\begin{equation}
\label{CP75}
\left( \omega_0 + \frac{3}{2} \right)\square\phi = 
\frac{T^{(m)}}{2} - 
2V(\phi) + \phi V^{\prime}(\phi) + \frac{\omega_{0}}{\phi} 
\left( \Pi^{2}
 - Q^{2} \right) \;. 
\end{equation}
Note that second derivatives of the scalar 
$\phi$ appear only in the form $\Box \phi$. 
For $\omega_0 = 0$ Brans-Dicke theory, equivalent to  metric $f
(R)$ gravity, one can use eq.~(\ref{CP75}) to eliminate completely the 
d'Alembertian $\Box \phi$ from the remaining equations. As a result,  
the Cauchy problem is well-formulated  in general and  well-posed 
in vacuo.  Further work by Salgado and 
collaborators \cite{Salgado2} 
established  the well-posedness of the Cauchy problem for scalar-tensor 
gravity with $\omega=1$ in the presence of matter, which implies  
 well-posedness for metric $f(R)$ gravity with 
matter along the lines established above.

Palatini $f(R)$ gravity, instead, is equivalent to an   
$\omega_0 = -3/2$  
Brans-Dicke theory: for this value of the Brans-Dicke parameter, the 
d'Alembertian $ \square\phi$ 
disappears from  eq.~(\ref{CP75}) and the field  $\phi$ is not 
dynamical (there is no wave equation to govern it): it can be specified 
arbitrarily  on a 
spacetime region provided that its gradient obeys the 
constraint~(\ref{CP75}). Hence, in  Palatini $f(R)$ gravity  it is 
impossible  
to  eliminate  $\square\phi$ from the 
system of differential equations unless, of course, $\square\phi 
= 
0$ (including the case of general relativity if $\phi =$ constant). 
Apart from the 
impossibility of 
a first-order formulation, one sees that for $\omega=-3/2$ the dynamical 
wave equation for $\phi$ is lost completely and this field is 
non-dynamical. Palatini  
$f(R)$ gravity has  an ill-formulated initial value problem even in 
vacuo and 
is regarded as a physically unviable theory.

An alternative approach to the initial value problem is by mapping the 
equivalent Brans-Dicke theory into its Einstein frame representation, in 
which the (redefined) scalar degree of freedom couples minimally to 
gravity but nonminimally to matter \cite{FaraoniTremblay}. This 
nonminimal coupling is, of 
course, absent in vacuo and, from the point 
of view of the Cauchy problem, plays a very minor role in the presence 
of  matter. In this approach, the non-dynamical role of the scalar is 
even  more obvious, and the conclusions above are reached by using 
well-known  theorems on the initial value problem of general relativity 
with a scalar field \cite{FaraoniTremblay}.

The problem with Palatini $f(R)$ gravity has been noticed with an 
entirely different approach, {\em  i.e.}, matching static interior and 
exterior solutions with spherical symmetry 
\cite{BarausseSotiriouMiller} (other problems are reported in 
Refs.~\cite{review,  Kaloper, PalatiniPLB}).

The field equations of Palatini $f(R)$ gravity  
are second order PDEs in the metric. Because  
$f$ is a function of $\tilde{R}$, which is 
an algebraic function of $T$ due to eq.~(\ref{Palatinitrace}), the right 
hand side of eq.~(\ref{Palatinireformulated}) includes second 
derivatives 
of $T$. But $T$ 
contains derivatives of the matter fields up to first order, hence    
eq.~(\ref{Palatinireformulated})  contains derivatives of the matter 
fields  
up to third order. This is in contrast with the situation of general 
relativity and  most of its extensions, in which the field 
equations contain only
first order derivatives of  the matter fields. As a consequence, in 
these theories the metric is generated by an integral 
over the 
matter sources and, therefore, discontinuities (or 
singularities) in the matter fields and their derivatives do not 
imply unphysical discontinuities  of the 
metric. In  
Palatini $f(R)$ gravity, instead, the algebraic dependence 
of  the metric on the matter fields creates unacceptable  
discontinuities  in the metric and singularities in the curvature, which 
is what is found in \cite{BarausseSotiriouMiller}. So, both the failure 
of the initial 
value problem and the occurrence of curvature singularities in the 
presence 
of discontinuities in the matter fields or their derivatives can be 
traced  to the fact that the scalar degree of freedom is 
non-dynamical and is related algebraically to $T$. A possible cure is to 
modify the gravitational sector of the action to raise the order 
of the field equations.

\section{Conclusions}
\setcounter{equation}{0}

We are now ready to summarize the situation of $f(R) $ gravity. Let us 
stress once again that we regard these theories more as toy models, and 
as  
proofs of principle that modified gravity can explain the observed 
acceleration of the universe without dark energy, than definitive 
theories. 

\begin{itemize}

\item {\bf Metric  $f(R)$ gravity:} models exist that pass all the 
observational and theoretical constraints. An example is 
the Starobinsky model \cite{Starobinsky07}
\be
f(R)=R+\lambda R_0 \left[ \frac{1}{\left( 1+ \frac{R^2}{R_0^2} \right)^n 
}-1 \right] \;.
\ee
All the viable models require the chameleon mechanism in order to pass 
the weak-field limit tests.
A condition that must be satisfied by all metric $f(R)$ theories 
in order to avoid the Dolgov-Kawasaki local instability is $f''(R) \geq 
0$. The condition~(\ref{stabilitydS})  must be satisfied for the 
stability of a de  Sitter space. The biggest problem is whether 
curvature singularities exist for relativistic strong field stars.

\item {\bf Palatini $f(R)$ gravity:} these theories  suffered multiple 
deaths; they contain a 
non-dynamical scalar field, the Cauchy problem is ill-posed, and 
discontinuities in the matter distribution generate curvature 
singularities.

\item {\bf Metric-affine gravity:} this class of theories  is not yet 
sufficiently developed to assess 
whether it is viable according to the criteria listed here, and its 
cosmological consequences are unexplored.

\end{itemize}

In conclusion, $f(R)$ theories have helped our understanding of the 
peculiarities of general relativity in the broader spectrum of 
relativistic theories of gravity, and have taught us about  
important aspects  
of its simple generalizations. They even constitute viable alternatives 
to dark energy models in explaining the cosmic acceleration, although at 
present there is no definite 
prediction that sets them apart once and for all from dark energy and 
other models.

\section*{Acknowledgments}

The author is grateful to Thomas Sotiriou for many discussions, to the 
International School 
for Adanced Studies in  Trieste, Italy, where this manuscript was 
prepared, for its hospitality, and to the Natural 
Sciences and  Engineering Research 
Council of Canada for financial support.

\end{document}